\documentclass{lmcs}
\usepackage{amsmath}
\usepackage{amssymb}

\usepackage{hyperref}

\usepackage{graphicx}
\usepackage{wasysym}
\usepackage{epstopdf}

\DeclareMathOperator{\M}{Max}
\DeclareMathOperator{\m}{Min}

\DeclareMathOperator{\val}{val}

\DeclareMathOperator{\reset}{dev}

\DeclareMathOperator{\bad}{Superf}
\DeclareMathOperator{\switch}{Switch_\sigma}
\newcommand{\prev}[2]{h_{#1}(#2)}
\newcommand{\tendvers}[1]{\mathop{\longrightarrow}_{#1}} %Olivier

\renewcommand{\paragraph}[1]{{\noindent\bf #1.}}

\newcommand{\NN}{\mathbb{N}}
\newcommand{\vertices}{V}
\newcommand{\vM}{\vertices_{\textrm{Max}}}
\newcommand{\vm}{\vertices_{\textrm{min}}}
\newcommand{\vr}{\vertices_{\textrm{R}}}

\newcommand{\sM}{\sigma}
\newcommand{\sm}{\tau}

\newcommand{\edges}{E}
\newcommand{\tp}[2]{p(#1|#2)}
\newcommand{\tpseul}{p}

\newcommand{\bh}{\setminus}

\newcommand{\fplay}{h}

\newcommand{\proba}[4]{\mathbb{P}^{#1,#2}_{#3}\left(#4\right)}
\newcommand{\probaa}[3]{\mathbb{P}^{#1,#2}_{#3}}
\newcommand{\esper}[4]{\mathbb{E}^{#1,#2}_{#3}\left[#4\right]}

\newcommand{\states}{V}
\newcommand{\wc}{W}

\newcommand{\taum}{\tau_{\frac{m}{2}}}

\newcommand{\un}{1\mskip-6.5mu 1}
\newcommand{\une}[1]{\un_{#1}}

\begin{document}

\title{Optimal Strategies in\\ Perfect-Information Stochastic Games\\
with Tail Winning Conditions}
\author[Gimbert]{Hugo Gimbert}	%required
\address{LaBRI, CNRS, Bordeaux, France}	%required
\email{hugo.gimbert@labri.fr}  %optional
%\thanks{thanks 1, optional.}	%optional

\author[Horn]{Florian Horn}	%optional
\address{LIAFA, Universit\'e Paris 7, Paris, France}	%optional
\email{florian.horn@liafa.jussieu.fr}  %optional
\thanks{This research was partially supported by french project ANR "DOTS".}	%optional

\subjclass{Games, Stochastic Processes}
%\titlecomment{}

\begin{abstract}
We prove that optimal strategies exist in perfect-information stochastic games
with finitely many states and actions and tail winning conditions.
\end{abstract}

\maketitle

\section*{Introduction}

We prove that optimal strategies exist in perfect-information stochastic games
with finitely many states and actions and tail winning conditions.

This proof is different from the algorithmic proof sketched in~\cite{theseflorian}.

\section{Perfect-Information Stochastic Games\label{sec:defs}}

In this section we give formal definitions of perfect-information stochastic games, values and
optimal strategies.

\subsection{Games, plays and strategies}
A (perfect-information stochastic) game is a tuple\\ $(\vertices, \vM, \vm, \vr,E,\wc,\tpseul)$,
where $(\vertices,\edges)$ is a finite graph, $(\vM,\vm,\vr)$ is a partition of
$\vertices$, $\wc \subseteq\vertices^\omega$ is a measurable set
called the winning condition and for
every $v\in\vr$ and $w\in\vertices$,
$\tp{w}{v}\geq 0$ is the transition probability from $v$ to $w$, with the
property $\sum_{w\in\vertices} \tp{w}{v} =1$.

A \emph{play} is an infinite sequence $v_0v_1\cdots\in\vertices^\omega$ of vertices such that if
$v_n\in(\vM\cup\vm)$ then $(v_n,v_{n+1})\in\edges$ and if $v_n\in\vr$
then $\tp{v_{n+1}}{v_n}>0$. 
A play is \emph{won by $\M$} if it belongs to $\wc$ otherwise the play is won by $\m$.
A \emph{finite play} is a finite prefix of a play.

A \emph{strategy} for player $\M$ is a mapping $\sM:\vertices^*\vM \to
\vertices$ such that for each finite play $\fplay = v_0 \ldots v_n$
such that $v_n \in \vM$, we have $(v_n,\sigma(\fplay))\in\edges$.
A play $v_0v_1\cdots$ is \emph{consistent with $\sM$} if
for every $n$, if $v_n\in\vM$ then $v_{n+1}$ is $\sigma(v_0 \cdots v_n)$. 
A strategy for player $\m$ is defined similarly, and is generally
denoted $\sm$.

Once the initial vertex $v$ and two strategies $\sM,\sm$ for player
$\M$ and $\m$ are fixed, we can measure the probability that a given set of
plays occurs. This probability measure is denoted
$\probaa{\sigma}{\tau}{v}$.
For every $n\in\NN$, we denote by $V_n$ the random variable defined by
$V_n(v_0v_1\cdots)=v_n$, the set of plays is equipped with the $\sigma$-algebra
generated by random variables $(V_n)_{n\in\NN}$. Then there exists a probability measure $\probaa{\sigma}{\tau}{v}$ with 
the following properties:
\begin{align}
&\proba{\sigma}{\tau}{v}{V_0=v}=1\\
\label{eq:vM}
&\proba{\sigma}{\tau}{v}{V_{n+1} = \sigma(V_0\cdots V_n)\mid V_n\in\vM}=1\enspace,\\
\label{eq:vm}
&\proba{\sigma}{\tau}{v}{V_{n+1} = \tau(V_0\cdots V_n)\mid V_n\in\vm}=1\enspace,\\
\label{eq:vr}
&\proba{\sigma}{\tau}{v}{V_{n+1}\mid V_n \in \vr} =
  \tp{V_{n+1}}{V_n}\enspace.
\end{align}

Expectation of a real-valued, measurable and bounded function $\phi$ under
$\probaa{\sigma}{\tau}{v}$ is denoted $\esper{\sigma}{\tau}{v}{\phi}$.
For an event $W\subseteq \states^\omega$, we denote $\une{W}$
the indicator function of $W$.
We will often use implicitely the following formula,
which gives the expectation of $\phi$ once a finite prefix $h=v_0v_1\cdots v_n$ of
the play is fixed:
\begin{equation}\label{eq:decal}
\esper{\sigma}{\tau}{v}{~\phi\mid V_0\cdots V_n = h}=
\esper{\sigma[h]}{\tau[h]}{v_n}{~\phi[h]~}\enspace,
\end{equation}
where $\sigma[h](w_0w_1w_2\cdots) = \sigma(v_0\cdots v_n w_1w_2\cdots)$ and
$\tau[h]$
and $\phi[h]$ are defined similarly.

\subsection{Values}
The goal of player $\M$ is to satisfy the winning condition
with the highest probability possible, whereas player $\m$ has the opposite
goal. Given a starting vertex $v$ and a strategy $\sigma$ for player
$\M$, whatever strategy $\tau$ is chosen by $\m$, the play will be won
with probability at least:
\[
\inf_{\tau} \proba{\sigma}{\tau}{v}{\wc}\enspace.
\]
Thus, starting from $v$, player $\M$ can ensure winning the game with
probability arbitrarily close to:
\[
\val_*(v) = \sup_{\sigma} \inf_{\tau}
\proba{\sigma}{\tau}{v}{\wc}\enspace,
\]
and symmetrically, player $\m$ can ensure the play is not won with
probability much higher
than:
\[
\val^*(v) =  \inf_{\tau}\sup_{\sigma}
\proba{\sigma}{\tau}{v}{\wc}\enspace.
\]
Clearly $\val_*(v) \leq \val^*(v)$.
According to Martin's theorem~\cite{martin} these values are equal, and 
this common value is called the value of vertex $v$ and denoted $\val(v)$

\subsection{Optimal and $\epsilon$-optimal strategies}
By definition of the value, for each $\epsilon>0$ there exist $\epsilon$-optimal strategies
$\sigma_\epsilon$ for player $\M$ and
$\tau_\epsilon$ for player $\m$ 
such that for every vertex $v$,
\[
 \inf_{\tau} \proba{\sigma_\epsilon}{\tau}{v}{\wc}\geq\val(v)-\epsilon\enspace,
\]
and symmetrically for player $2$,\[
 \sup_{\sigma} \proba{\sigma}{\tau_\epsilon}{v}{\wc}\leq\val(v)+\epsilon\enspace.
\]

For several classes of winning conditions, it is known that there exists \emph{optimal strategies},
i.e. strategies that are $\epsilon$-optimal for every $\epsilon$.

In this paper,
we prove that optimal strategies exist in games whose winning condition has the following property.
\begin{defi}
A winning condition $\wc \subseteq V^\omega$ is a \emph{tail winning condition} if for every finite play $p\in V^*$
and infinite play $q\in V^\omega$,
\[
(q \in \wc) \iff (pq \in\wc)\enspace.
\]
\end{defi}

Games with tail winning conditions have the following properties.
\begin{lem}\label{lem:eq}
Let $G$ be a game with a tail winning condition $\wc$.
Then for every vertex $v\in V$,
\[
\begin{cases}
\val(v)=\max_{(v,w)\in E} \val(w) & \text{ if } v\in\vM\enspace,\\
\val(v)=\min_{(v,w)\in E} \val(w) & \text{ if } v\in\vm\enspace,\\
\val(v)=\sum_{(v,w)\in E} \tp{w}{v} \val(w)& \text{ if } v\in\vr\enspace.
\end{cases}
\]
\end{lem}

\begin{proof}
This comes from~\eqref{eq:decal}, and the fact that $\une{\wc}[h]=\une{\wc}$,
because $\wc$ is a tail winning condition.
\end{proof}

\section{Optimal strategies in games with tail winning conditions}

Our main result is:

\begin{thm}\label{theo:main}
In every perfect-information stochastic game with tail winning condition
and finitely many states and actions,
both players have optimal strategies.
\end{thm}

The proof of this theorem relies on several intermediary results.

%%%%%%%%%%%%%%%%%%%%%%%%%%%%%%%%%%%%%%%%%%%%%%
%%%%%%%%%%%%%%%consistent games
%%%%%%%%%%%%%%%%%%%%%%%%%%%%%%%%%%%%%%%%%%%%

\subsection{Consistent games}

Next lemma states that it is enough to prove Theorem~\ref{theo:main}
in the case where no move of player $\M$ can decrease the value of a vertex
and no move of player $\m$ can increase the value of a vertex.
%i.e. when inequalities of
%Lemma~\ref{lem:eq} are equalities.

\begin{lem}\label{lem:delete}
Let $G$ be a game with a tail winning condition $\wc$.
We say an edge $(v,w)$ is superfluous when either
$v\in\vM$ and $\val_G(w)<\val_G(v)$ or $v\in\vm$ and $\val_G(w)>\val_G(v)$.
Let $G'$ the game obtained from $G$ by removing all superfluous edges.
If there are optimal strategies in $G'$ then there are optimal strategies in $G$ as well.
\end{lem}
\begin{proof}
We prove that there exists optimal strategies
in the game $G'$ obtained by removing
only one of the superfluous edges,
Lemma~\ref{lem:delete} then results from a trivial induction.

Let $(v_s,w_s)$ be the superfluous edge removed.
Without loss
of generality, suppose $v_s\in\vM$,
and let
\[
m=\val_G(v_s)-\val_G(w_s)>0\enspace.
\]

Suppose there exists optimal strategies $\sigma',\tau'$ in $G'$.

In game $G$, player $\M$ has more freedom than in game $G'$,
and from every vertex $v$
player $\M$ can guarantee the probability to win to be
at least $\val_{G'}(v)$, for that player $\M$ can use its strategy $\sigma'$
for $G'$, which is a strategy in $G$ as well.

We are going to show that this is the best that player $\M$
can expect in $G$: we are going
to build a strategy $\tau$ that prevents the probability to win to be
greater than $\val_{G'}$.
As a consequence, $\sigma'$ and $\tau$ are a couple of optimal strategies in $G$,
which proves the lemma.

The strategy $\tau$ is as follows.
As long as player $\M$ does not choose the superfluous edge $(v_s,w_s)$,
the play is a play in $G'$ and strategy $\tau$ consists in playing like the strategy $\tau'$ in $G'$.
If at some moment player $\M$ chooses the superfluous edge $(v_s,w_s)$
then strategy $\tau$ forgets the prefix of the play and switches definitively to a $\frac{m}{2}$-optimal
strategy $\taum$ in $G$.  If subsequently player $\M$ chooses the superfluous edge again,
nothing special happens, $\tau$ keeps playing accordingly to $\tau'$.
Let $\bad$ be the event defined by:
\[
\bad=\{ \exists n \in \NN, (V_n,V_{n+1})=(v_s,w_s)\}\enspace,
\]
then the definition of $\tau$ and $m$ ensures that for any strategy $\sigma$ and vertex $v$,
%By definition, strategy $\tau$ switches to a $\frac{m}{2}$-optimal strategy as soon as the event $\dev_\epsilon$ occurs,
%hence $
\begin{equation}\label{eq:stepp0}
\proba{\sigma}{\tau}{v}{\wc\mid \bad}\leq\val_G(w_s)+\frac{m}{2} = \val_G(v_s)-\frac{m}{2} \enspace.
%$.
\end{equation}

That way we have an upper bound on the probability to win when the plays does go through the superfluous edge.
In case the play does \emph{not} go through the superfluous edge, we prove:
%About plays that do not use the superfluous edge, we are going to prove that for every strategy $\sigma$,
\begin{equation}\label{eq:stepp1}
\proba{\sigma}{\tau}{v_s}{\wc\mid\neg\bad}
\leq
 \val_{G'}(v_s)\enspace.
\end{equation}
For this, we use the following transformation of $\sigma$ into a strategy $\sigma_s$
in $G'$.
Strategy $\sigma_s$ plays similarly to $\sigma$ as long as strategy $\sigma$ does not
plays the superfluous edge $(v_s,w_s)$.
If after a finite play $v_0,\ldots,v_n$, with $v_n=v_s$, strategy $\sigma$ is about to choose the superfluous edge $(v_s,w_s)$,
then $\sigma_s$ stops playing similarly to $\sigma$.
Instead, strategy  $\sigma_s$ forgets the past and switches definitively to the strategy $\sigma'$ optimal in $G'$,
in other words for every play $p$, $\sigma_s(v_0\cdots v_n p)=\sigma'(p)$.
We denote $\switch$ the event:
\[
\switch = \{\exists n\in\NN, V_n=v_s \text{ and }\sigma(V_0,\ldots,V_n)=w_s \}\enspace.
\]
Then by definition of $\sigma_s$, for every strategy $\sigma$ and vertex $v$,
\begin{align}
\label{eq:coincide}
&\proba{\sigma_s}{\tau}{v}{\wc\mid\neg\switch}
=\proba{\sigma}{\tau}{v}{\wc\mid\neg\bad}\\
\label{eq:switch}
&\proba{\sigma_s}{\tau}{v}{\wc\mid\switch}\geq\val_{G'}(v_s)\enspace.
\end{align}
Since $\sigma_s$ is a strategy in $G'$ then
$\proba{\sigma_s}{\tau}{v}{\wc}
=\proba{\sigma_s}{\tau'}{v}{\wc}\leq\val(G')(v_s)$
because $\tau'$ is optimal in $G'$.
Since $\proba{\sigma_s}{\tau}{v}{\wc}$
is a convex combination of
$\proba{\sigma_s}{\tau}{v}{\wc\mid\neg\switch}$
and
$\proba{\sigma_s}{\tau}{v}{\wc\mid\switch}$
then according to~\eqref{eq:switch} it implies that
$\proba{\sigma_s}{\tau}{v}{\wc\mid\neg\switch}\leq\val_{G'}(v_s)$.
Together with~\eqref{eq:coincide} it proves~\eqref{eq:stepp1}.

We can now prove that the value of $v_s$ in $G$ and $G'$ are the same:
\begin{equation}\label{eq:thesame}
\val_G(v_s)=\val_{G'}(v_s)\enspace.
\end{equation}
Indeed, for every strategy $\sigma$,
$\proba{\sigma_s}{\tau}{v}{\wc}$ is a convex combination
of
$\proba{\sigma_s}{\tau}{v}{\wc\mid \bad}$
and
$\proba{\sigma_s}{\tau}{v}{\wc\mid \neg\bad}$
hence according to~\eqref{eq:stepp0} and~\eqref{eq:stepp1},
$\proba{\sigma_s}{\tau}{v}{\wc}\leq \max\{\val_G{v_s}-\frac{m}{2},\val_{G'}(v_s)\}$.
Taking the supremum over $\sigma$, since $m>0$ it proves~\eqref{eq:thesame}.

To conclude we prove that~\eqref{eq:thesame} holds not only for $v_s$
but for any vertex $v$. Let $v$ be a vertex,
$\sigma$ be a strategy and $\sigma_s$ the associated switch strategy.
Then, since $\sigma$ and $\sigma_s$ coincide when
event $\bad$ does not occur,
\begin{align}
\notag
\proba{\sigma}{\tau}{v}{\wc}&=
\proba{\sigma}{\tau}{v}{\wc\land\neg\bad}+
\proba{\sigma}{\tau}{v}{\wc\mid\bad}
\cdot
\proba{\sigma}{\tau}{v}{\bad}
\\
\label{eq:ouf}
&=
\proba{\sigma_s}{\tau}{v}{\wc\land\neg\bad}+
\proba{\sigma}{\tau}{v}{\wc\mid\bad}
\cdot
\proba{\sigma_s}{\tau}{v}{\bad}\enspace.
\end{align}
According to~\eqref{eq:stepp0},
$\proba{\sigma}{\tau}{v}{\wc\mid\bad}\leq\val_G(v_s)
=\val_{G'}(v_s)$ according to~\eqref{eq:thesame}.
By definition of $\tau$ and $\sigma_s$,
$\proba{\sigma_s}{\tau}{v}{\wc\mid\bad}=\val_{G'}(v_s)$
because when the event $\bad$ occurs the play is consistent with
optimal strategies $\sigma'$ and $\tau'$ in $G'$.
Finally, $\proba{\sigma}{\tau}{v}{\wc\mid\bad}
\leq
\proba{\sigma_s}{\tau}{v}{\wc\mid\bad}$,
which together with~\eqref{eq:ouf}
gives
$\proba{\sigma}{\tau}{v}{\wc}\leq
\proba{\sigma_s}{\tau}{v}{\wc}$.
Since $\sigma_s$ is a strategy in $G'$ and $\tau$ is optimal in $G'$,
$\proba{\sigma_s}{\tau}{v}{\wc}\leq\val_{G'}(v)$.
Taking the supremum over $\sigma$,
we get $\val_G(v)\leq\val_{G'}(v)$ which achieves the proof.
\end{proof}

We say that a game $G$ is \emph{consistent} when for every edge $(v,w)$,
if $v\in \vM\cup\vm$ then $\val_G(v)=\val_G(w)$.
consistent games have the following properties.
\begin{lem}\label{lem:consistent}
Let $G$ be a consistent game with a tail winning condition $\wc$.
Then for every initial vertex $v_0$ and strategies $\sigma,\tau$, and every $n\in\NN$,
\[
\esper{\sigma}{\tau}{v_0}{\val(V_{n+1})\mid V_0,\ldots,V_{n}} = \val(V_{n})\enspace.
\]
\end{lem}
\begin{proof}
Comes from Lemma~\ref{lem:eq} and the fact that the game is consistent.
\end{proof}

%%%%%%%%%%%%%%%%%%%%%%%%%%%%%%%%%%%%%%%%%%%%%%
%%%%%%%%%%%%%%%               Deviations
%%%%%%%%%%%%%%%%%%%%%%%%%%%%%%%%%%%%%%%%%%%%

\subsection{Deviations}

To detect bad behaviours of a strategy,
we use the notions of quality and deviations.

%\begin{defi}[Quality and deviations of a strategy]
The \emph{quality of a strategy} $\sigma$ after a finite play is
\[
\prev{\sigma}{v_0,\ldots,v_n} = \inf_\tau \proba{\sigma}{\tau}{v}{\wc\mid V_0=v_0,\ldots,V_n=v_n}\enspace.
\]
%To detect when $\sigma$ starts playing worse than a

A \emph{deviation} occurs when the quality of the strategy drops significantly below the value of the current vertex.
Formally,
%$\sigma$ is a finite play from which $\sigma$
%starts playing less than $\frac{m}{2}$-optimally,
%where
%Moreover, for detecting when $\sigma$ starts playing bad
%we define 
let
\[
m=\min_{v\in V}\{\val(v),  \val(v)>0\}\enspace,
\]
be the smallest strictly positive value\footnote{if $\forall v\in V, \val(v)=0$ then $m=\infty$ however this case has no interest.}
 of a vertex in $G$,
the \emph{deviation date} is denoted $\reset_\sigma$ and defined by:
\[
\reset_\sigma = \min \left\{n\mid
\prev{\sigma}{V_0,\ldots,V_n} \leq \val(V_n)-\frac{m}{2}
\right\}\enspace,
\]
with the convention $\min \emptyset =\infty$.
%\end{defi}

Next lemma states that when player $\M$ plays $\epsilon$-optimally,
with $\epsilon$ small enough,
deviations occur with probability strictly less than $1$.

\begin{lem}\label{lem:finish}
Let $G$ be a consistent game with a tail winning condition $\wc$.
Let $\epsilon>0$ and $\sigma$ be an $\epsilon$-optimal strategy.
For every vertex $v$ and strategy $\tau$,
\begin{equation}
\label{eq:dev2}
\proba{\sigma}{\tau}{v_0}{\reset_\sigma<\infty } \leq \frac{1%\frac{m}{2}
 + \epsilon}{1+\frac{m}{2}}\enspace.
\end{equation}
\end{lem}

\begin{proof}
We start the proof with a modification of $\tau$ and introduce an auxiliary strategy $\tau'$.
with the following properties:
\begin{equation}
\label{eq:tauliketauprime}
\proba{\sigma}{\tau'}{v_0}{\reset_\sigma<\infty }  = 
\proba{\sigma}{\tau}{v_0}{\reset_\sigma<\infty }
\enspace.
\end{equation}
Let $\epsilon'>0$.
Strategy $\tau'$ plays like strategy $\tau$ as long as there is no deviation
i.e. as long as $\prev{\sigma}{v_0,\ldots,v_n} > \val(v_n)-\frac{m}{2}$.
In case a deviation occurs i.e.
$\prev{\sigma}{v_0,\ldots,v_n} \leq \val(v_n)-\frac{m}{2}$
then strategy $\tau'$ forgets the past and switches definitively
to an $\epsilon'$-optimal response to
$\sigma[v_0,\ldots,v_n]$, so that
\begin{equation}\label{eq:concl3}
\proba{\sigma}{\tau'}{v_0}{W\mid \reset_\sigma =n \text{ and } V_0\cdots V_n=v_0\cdots v_n}\leq \val(v_n)-\frac{m}{2}+\epsilon'\enspace.
\end{equation}
The equality~\eqref{eq:tauliketauprime} holds because $\tau$ and $\tau'$ coincide as long as there is no deviation.

We start with proving:
% for every $n\in\NN$,
\begin{equation}
\label{eq:dev1}
\esper{\sigma}{\tau'}{v_0}{\val\left(V_{\reset_\sigma}\right)\cdot\une{\reset_\sigma<\infty}}\leq \val(v_0)\enspace.
\end{equation}
For every $n\in\NN$ let $\reset_\sigma^{(n)}=\min\{n,\reset_\sigma\}$.
According to Lemma~\ref{lem:consistent},
$\esper{\sigma}{\tau'}{v_0}{\val\left(V_{\reset_\sigma^{(n)}}\right)} = \val(v_0)$
hence
$\esper{\sigma}{\tau'}{v_0}{\val\left(V_{\reset_\sigma^{(n)}}\right)\cdot\une{\reset_\sigma<n}} \leq \val(v_0)$.
Taking the limit of the left hand-side of this inequality when $n\to \infty$, we obtain~\eqref{eq:dev1}.

The main step of the proof is to establish:
\begin{equation}\label{eq:concl2}
\proba{\sigma}{\tau'}{v_0}{W\land \reset_\sigma <\infty}\leq \val(v_0)-\frac{m}{2}\cdot \proba{\sigma}{\tau'}{v_0}{\reset_\sigma <\infty}\enspace.
\end{equation}
Then,
\begin{align*}
\proba{\sigma}{\tau'}{v_0}{W\land \reset_\sigma <\infty}&
=
\esper{\sigma}{\tau'}{v_0}{\une{W}\cdot \une{\reset_\sigma <\infty}}\\
&=
\esper{\sigma}{\tau'}{v_0}{
\esper{\sigma}{\tau'}{v_0}{\une{W}\cdot \une{\reset_\sigma <\infty}\mid \reset_\sigma,V_0,\ldots,V_{\reset_\sigma}}
}\\
&=
\esper{\sigma}{\tau'}{v_0}{
\esper{\sigma}{\tau'}{v_0}{\une{W}\mid \reset_\sigma,V_0,\ldots,V_{\reset_\sigma}}\cdot \une{\reset_\sigma <\infty}
}\\
&\leq
\esper{\sigma}{\tau'}{v_0}{
\left(\val(V_{\reset_\sigma}) -\frac{m}{2} +\epsilon' \right)\cdot \une{\reset_\sigma <\infty}
}\\
&=
\esper{\sigma}{\tau'}{v_0}{
\val(V_{\reset_\sigma})\cdot \une{\reset_\sigma <\infty}
} + \left(-\frac{m}{2} +\epsilon'\right)\cdot
\proba{\sigma}{\tau'}{v_0}{\reset_\sigma <\infty}
 \\
&\leq
\val(v_0)+
\left(-\frac{m}{2} +\epsilon'\right)\cdot\proba{\sigma}{\tau'}{v_0}{\reset_\sigma <\infty}
\enspace,
\end{align*}
where the three first equalities are properties of conditional expectations,
the first inequality is~\eqref{eq:concl3}
and the second inequality is~\eqref{eq:dev1}.
Since this holds for every $\epsilon'$, we obtain~\eqref{eq:concl2} as promised.

Now we can conclude.
Since $\sigma$ is $\epsilon$-optimal,
\begin{align}
\val(v_0)-\epsilon
&
\notag
\leq\proba{\sigma}{\tau'}{v_0}{W}
=
\proba{\sigma}{\tau'}{v_0}{W\land \reset_\sigma <\infty}
+
\notag
\proba{\sigma}{\tau'}{v_0}{W\land\reset_\sigma =\infty}\\
&\leq
\proba{\sigma}{\tau'}{v_0}{W\land \reset_\sigma <\infty}
+
\label{eq:concl}
1 - \proba{\sigma}{\tau'}{v_0}{\reset_\sigma<\infty}\enspace.
\end{align}
Together with~\eqref{eq:concl2} we obtain~\eqref{eq:dev2}
with $\tau'$ instead of $\tau$ and according to~\eqref{eq:tauliketauprime}
this completes the proof of the lemma.
\end{proof}

\subsection{Construction of an optimal strategy}
We can now proceed with the second and last step in the proof of Theorem~\ref{theo:main}.
From an $\epsilon$-optimal strategy $\sigma$, with $\epsilon$ small enough,
we construct an optimal strategy, by resetting the memory of $\sigma$ at right moments.
A similar construction has been used in~\cite{tail} for proving a zero--one law
in concurrent games with tail winning conditions.

%The quality of a strategy is closely linked to the probability to win:
%\begin{lem}\label{lem:ultimate}
%For every initial vertex $v_0$ and strategies $\sigma,\tau$,
%\[
%\proba{\sigma}{\tau}{v_0}{\prev{\sigma}{V_0,\ldots,V_n} \text{ converges to }\une{\wc}}=1\enspace.
%%=
%%\limsup_n
%%\proba{\sigma}{\tau}{v_0}{\limsup_n \prev{\sigma}{V_0,\ldots,V_n}>0}\enspace.
%%\esper{\sigma}{\tau}{v_0}{\prev{\sigma}{V_0,\ldots,V_n}}\enspace.
%\]
%\end{lem}
%\begin{proof}
%This is Levy's or Kolmogorov's law, see~\cite{durett}. 
%\end{proof}

\begin{lem}\label{lem:optnorm}
Let $G$ be a consistent game with a tail winning condition $\wc$.
Then player $\M$ has an optimal strategy in $G$.
\end{lem}

\begin{proof}
If all vertices in $G$ have value $0$,
there is nothing to prove.

Otherwise, let $m$ be the smallest strictly positive value of a vertex
and $\sigma$ be an $\frac{m}{4}$-optimal strategy.
Using $\sigma$, we are going to define a strategy $\sigma'$ and prove that $\sigma'$ is optimal in $G$.
For that, we define $t(v_0,\ldots,v_n)$ the date of the latest deviation before date $n$ by $t(v_0)=0$ and
%\begin{multline}
\[
t(v_0,\ldots,v_n,v_{n+1})=
\begin{cases}
t(v_0,\ldots,v_n) &\text{ if } \prev{\sigma}{v_{t(v_0,\ldots,v_n)},\ldots,v_{n+1}} \geq \val(v_{n+1})-\frac{m}{2}\enspace,\\
n+1& \text{ otherwise}.
\end{cases}
\]
%\end{multline}
%the set in this definition always contains $n$ because $\sigma$ is $\frac{m}{4}$-optimal.
By definition the sequence $(t(V_0,\ldots,V_n))_{n\in\NN}$ is increasing,
we denote $T$ its limit in $\NN \cup \{\infty\}$.
Strategy $\sigma'$ consists in forgetting everything before the last deviation and applying $\sigma$,
i.e.
\[
\sigma'(v_0,\ldots,v_n)=\sigma(v_{t(v_0,\ldots,v_n)},\ldots,v_n)\enspace.
\]

%\medskip

To prove that $\sigma'$ is optimal, we start with proving for every strategy $\tau$ and vertex $v$,
\begin{equation}\label{eq:finite}
\proba{\sigma'}{\tau}{v}{T < \infty}=1
\enspace.
\end{equation}
Let $D=\min\{n\mid t(V_0,\ldots,V_n)\geq 1\}$ be the date of the first deviation,
then since $\sigma$ and $\sigma'$ coincide until the first deviation,

\begin{equation}
\label{eq:co1}
\proba{\sigma'}{\tau}{v}{D<\infty}=\proba{\sigma}{\tau}{v}{D<\infty}\enspace,
\end{equation}
and by definition of $\sigma'$ for every $n\in\NN$,
\begin{equation}
\label{eq:co2}
\proba{\sigma'}{\tau}{v}{T=\infty\mid D=n, V_0=v_0,\ldots,V_n=v_n}=\proba{\sigma'}{\tau[v_0,\ldots,v_n]}{v_n}{T=\infty}\enspace.
\end{equation}
%We denote by $T$ the possibly infinite date of the last deviation,
Let $\epsilon>0$ and $\tau$ and $v$ such that:
\begin{equation}\label{eq:deftauv}
\sup_{\tau',v'} \proba{\sigma'}{\tau'}{v'}{T=\infty}\leq \proba{\sigma'}{\tau}{v}{T=\infty} +\epsilon\enspace.
%\enspace.
\end{equation}
% such that $\proba{\sigma'}{\tau}{v}{T=\infty}>0$
%and let seek a contradiction.
%then
%\begin{equation}\label{eq:firstdev}
%$(D<\infty)\iff (T\geq1)$.
According to lemma~\ref{lem:finish},
since $\sigma$ is $\frac{m}{4}$-optimal,
%this gives
%the probability that at least one deviation occurs is
\begin{equation}\label{eq:firstdev}
%\forall \tau,\forall v,
 \proba{\sigma}{\tau}{v}{D<\infty}\leq \frac{1+\frac{m}{4}}{1+\frac{m}{2}}  <1
\enspace.
\end{equation}
By properties of conditional expectations,
\begin{align*}
\proba{\sigma'}{\tau}{v}{T=\infty}
&=
\esper{\sigma'}{\tau}{v}{
\proba{\sigma'}{\tau}{v}{T=\infty\mid D, V_0,\ldots, V_D}}\\
&=
\esper{\sigma'}{\tau}{v}{
\une{D<\infty}\cdot\proba{\sigma'}{\tau}{v}{T=\infty\mid D, V_0,\ldots, V_D}}\\
&=
\esper{\sigma'}{\tau}{v}{
\une{D<\infty}\cdot\proba{\sigma'}{\tau[V_0,\ldots,V_D]}{V_D}{T=\infty}}\\
&\leq
\esper{\sigma'}{\tau}{v}{
\une{D<\infty}\cdot\left(\proba{\sigma'}{\tau}{v}{T=\infty}+\epsilon\right)
}
\\
&=
\proba{\sigma'}{\tau}{v}{D<\infty}\cdot \left(\proba{\sigma'}{\tau}{v}{T=\infty}+\epsilon\right)
\\
%&=
%\proba{\sigma}{\tau}{v}{D<\infty}\cdot \left(\proba{\sigma'}{\tau}{v}{T=\infty}+\epsilon\right)
%\\
&=
\frac{1+\frac{m}{4}}{1+\frac{m}{2}}
\cdot \left(\proba{\sigma'}{\tau}{v}{T=\infty}+\epsilon\right)\enspace,
\end{align*}
where
the second equality is because $\proba{\sigma}{\tau}{v}{D<\infty\mid T=\infty}=1$,
the third equality is~\eqref{eq:co2},
the inequality is~\eqref{eq:deftauv},
and the last equality is~\eqref{eq:co1} and~\eqref{eq:firstdev}.
%Together with~\eqref{eq:deftauv} and~\eqref{eq:firstdev},
%this implies 
%\[
%\sup_{\tau',v'} \proba{\sigma'}{\tau'}{v'}{T=\infty}\leq \frac{\epsilon}{1-\proba{\sigma}{\tau}{v}{D<\infty}}\enspace.
%\]
Since this holds for any $\epsilon$, we obtain $\proba{\sigma'}{\tau}{v}{T=\infty}=0$
i.e.~\eqref{eq:finite}.

Second step of the proof is to establish:
\begin{equation}\label{eq:ffinite}
\proba{\sigma'}{\tau}{v_0}{\val(V_n)\tendvers{n\to\infty} 0
\mid
\prev{\sigma'}{V_0,\ldots,V_n}\tendvers{n\to\infty} 0}
=1
%\leq
%\proba{\sigma'}{\tau}{v_0}{
%}
%\proba{\sigma'}{\tau}{v_0}{\prev{\sigma'}{V_0,\ldots,V_n}\tendvers{n\to\infty} 0}\leq
%\proba{\sigma'}{\tau}{v_0}{\val(V_n)\tendvers{n\to\infty} 0}
\enspace.
\end{equation}
When playing with $\sigma'$,
suppose $\prev{\sigma'}{V_0,\ldots,V_n}$ converges to $0$
then by definition of $\sigma'$,
$\prev{\sigma}{V_{t(V_0,\ldots,V_n)},\ldots,V_n}$ converges to $0$ as well.
According to~\eqref{eq:finite}, $t(V_0,\ldots,V_n)$ has limit $T<\infty$ hence
$\prev{\sigma}{V_{T},\ldots,V_n}$ converges to $0$ as well.
By definition of $T$, for every $n\geq T$, $t(V_0,\ldots,v_n)=T$,
hence $\prev{\sigma}{V_{T},\ldots,V_n}\geq \val(V_n)-\frac{m}{2}$.
Since $\prev{\sigma}{V_{T},\ldots,V_n}$ converges to $0$,
$\limsup_n \val(V_n)\leq \frac{m}{2}$.
But 
hence
$\val(V_n)\tendvers{n} 0$ because by definition of $m$,
$(\val(v) <m)\implies(\val(v)=0)$.
This proves~\eqref{eq:ffinite}.

We can now achieve the proof of the optimality of $\sigma'$.
Since $\wc$ is a tail winning condition,
Levy's law~\cite{durett} implies,
\begin{align*}
\proba{\sigma'}{\tau}{v_0}{\neg\wc}
&= \proba{\sigma'}{\tau}{v_0}{
\proba{\sigma'}{\tau}{v_0}{W\mid V_0,\ldots,V_n}
\tendvers{n} 0}\\
&\leq \proba{\sigma'}{\tau}{v_0}{\prev{\sigma'}{V_0,\ldots,V_n}\tendvers{n} 0}\\
&\leq
\proba{\sigma'}{\tau}{v_0}{\val(V_n)\tendvers{n} 0}\\
&\leq \esper{\sigma'}{\tau}{v_0}{1 - \limsup_n \val(V_n)}\\
&\leq 1 - \limsup_n \esper{\sigma'}{\tau}{v_0}{\val(V_n)}\\
&=1-\val(v_0)\enspace,
\end{align*}
where the first inequality holds by definition of $\prev{\sigma'}{v_0,\ldots,v_n}$,
the second is~\eqref{eq:ffinite}, the third and fourth are basic properties of expectation
and the last equality holds according to lemma~\ref{lem:consistent}.
This proves that $\sigma'$ is optimal in $G$.
\end{proof}
\subsection{Proof of Theorem~\ref{theo:main}}
According to lemma~\ref{lem:delete} we can suppose without loss of generality that $G$ is consistent.
Since both the winning condition $\wc$ and its complement $\states^\omega \bh \wc$ are tail
winning conditions,
lemma~\ref{lem:optnorm} implies that both players have optimal strategies in $G$.

\section*{Conclusion}
We have proved the existence of optimal strategies in any perfect-information game with
a tail winning condition.
We relied heavily on the finiteness of the game,
actually the result does not hold in general for infinite arenas.
Extension of this result to certain classes of games with partial information
or with infinitely many vertices seems to be an interesting research direction.

\bibliographystyle{alpha}
\bibliography{optimal}
\end{document}